\documentclass[reprint,aps,pra,showpacs,floatfix,twocolumn]{revtex4-1}
\usepackage{graphicx}
\usepackage{dsfont}
\usepackage{epstopdf}

\newcommand{\be}{\begin{equation}}
\newcommand{\ee}{\end{equation}}

\begin{document}

\title{Limit distributions of three-state quantum walks: the role of coin eigenstates}

\author{M. \v Stefa\v n\'ak\email[correspondence to:]{martin.stefanak@fjfi.cvut.cz}}
\affiliation{Department of Physics, Faculty of Nuclear Sciences and Physical Engineering, Czech Technical University in Prague, B\v
rehov\'a 7, 115 19 Praha 1 - Star\'e M\v{e}sto, Czech Republic}

\author{I. Bezd\v ekov\'a}
\affiliation{Department of Physics, Faculty of Nuclear Sciences and Physical Engineering, Czech Technical University in Prague, B\v
rehov\'a 7, 115 19 Praha 1 - Star\'e M\v{e}sto, Czech Republic}

\author{I. Jex}
\affiliation{Department of Physics, Faculty of Nuclear Sciences and Physical Engineering, Czech Technical University in Prague, B\v
rehov\'a 7, 115 19 Praha 1 - Star\'e M\v{e}sto, Czech Republic}


\date{\today}

\begin{abstract}
We analyze two families of three-state quantum walks which show the localization effect. We focus on the role of the initial coin state and its coherence in controlling the properties of the quantum walk. In particular, we show that the description of the walk simplifies considerably when the initial coin state is decomposed in the basis formed by the eigenvectors of the coin operator. This allows us to express the limit distributions in a much more convenient form. Consequently, striking features which are hidden in the standard basis description are easily identified. Moreover, the dependence of moments of the position distribution on the initial coin state can be analyzed in full detail. In particular, we find that in the eigenvector basis the even moments and the localization probability at the origin depend only on incoherent combination of probabilities. In contrast, odd moments and localization outside the origin are affected by the coherence of the initial coin state.
\end{abstract}

\maketitle

\section{Introduction}
\label{sec1}

Quantum walks \cite{adz,meyer,fg} emerged as an extension of a concept of a random walk to a unitary evolution of a quantum particle on a discrete graph or lattice. Soon, their potential for quantum information processing was recognized \cite{kempe}. The quantum walk based algorithms which outperforms their classical counterparts were implemented for problems such as database search \cite{skw} or finding a path in a randomly glued tree graphs \cite{childs:tree}. Later it was shown \cite{childs,Lovett} that quantum walks represent a universal tool for quantum computation. More recently quantum walk based algorithms are being developed for problems such as graph isomorphism testing \cite{gamble,berry,rudiger} or finding structural anomalies in graphs \cite{reitzner,hillery,cottrell}.

For quite a long time the quantum walk was rather a theoretical concept, even though very fruitful. However, in 2009 the first experimental realization of a quantum walk on a line utilizing optically trapped atoms \cite{karski} has been reported. The experiments with cold ions \cite{schmitz,Zahringer} and photons \cite{Schreiber,Broome} followed shortly afterwards. These experiments were latter expanded to implement a quantum walk on a line with two non-interacting particles \cite{peruzzo:corelated:photons,two:photon:waveguide,sansoni}. More recently, an experiment implementing a quantum walk on a square lattice \cite{and:2dwalk:science} has been realized which was capable of simulating the walk on a line of two interacting particles \cite{ahlbrecht:int}.

One of the distinctive features of quantum walks when compared with classical random walks is their quadratically faster spreading. This stems from the fact that the quantum walk is a wave phenomenon \cite{knight} rather than a diffusion. The probability distributions resulting from quantum walks have typically an inverted-bell shape with characteristic peaks on the edges which propagate through the lattice with constant velocity. For quantum walks with homogeneous coin the probability distribution can be investigated by means of Fourier analysis \cite{ambainis}. Important results are the weak-limit theorems \cite{Grimmett} which prove the convergence of the moments of the re-scaled position of the quantum particle in the limit of large number of steps. This allows one to derive the so-called group-velocity density which can be used to approximate the probability distribution generated by the quantum walk and to evaluate the moments of this distribution. More recently, a method based on matrix-valued orthogonal polynomials have been developed \cite{cgmv} which extends the analysis to models with inhomogeneities.

In the present paper we investigate the position distributions of two one-parameter families of three-state quantum walks which we have introduced in \cite{stef:cont:def}. These families of quantum walks were derived as extensions of the three-state Grover walk. This particular model was extensively studied in the literature \cite{inui:pre,inui:psa} where it was found that it differs considerably from the two-state walk \cite{konno:limit:2002}. Namely, the three-state Grover walk features the so-called localization effect, which means that the particle has a non-vanishing probability to stay at any position even in the limit of infinite number of steps. The reason why does this effect appear is that the evolution operator of the three-state Grover walk possess apart from continuous spectrum also an isolated eigenvalue. The corresponding stationary state is unfolded over the whole lattice. Provided that it has a non-zero overlap with the initial condition, part of the wave-packet describing the quantum particle will remain trapped at the already visited sites and will not evolve anymore. This results in an additional central peak in the probability distribution of the three-state Grover walk which is exponentially decaying with the distance from the origin but is independent of the number of steps. The same holds for the two families of quantum walks we have derived in \cite{stef:cont:def} since they preserve the point spectrum. However, they do not exhaust the set of quantum walks with localization, as we have recently shown in \cite{stef:3state}. The probability distribution of such quantum walks is not described solely by the group-velocity density, since this part takes into account only the continuous spectrum. To obtain the full probability distribution one has to include also the localization stemming from the point spectrum. For the three-state Grover walk these two parts of the probability distribution were derived recently by Falkner and Boettcher in \cite{falkner}, where the authors have also discussed the rate of convergence of the moments of the distribution to the asymptotic values determined by the group-velocity density. At the same time, Machida \cite{machida} presented similar results extended to one of the families of quantum walks with localization we have introduced in \cite{stef:cont:def} and discussed the application of the three-state walk for preparation of discrete uniform measures. However, both results \cite{falkner,machida} have a significant drawback. Namely, the dependence of both the group-velocity density and the localization probability on the initial coin state is rather involved. The reason for this inconvenience is that the initial coin state is expressed in the standard basis of the coin space. However, we can describe the initial state in a different basis which proves to be more suitable for the analysis of the quantum walk. In \cite{stef:dir:col} we have shown that for the Hadamard walk on a line this suitable basis is given by the eigenstates of the coin operator. The approach resembles the transformation from bare states to dressed states familiar from quantum optics \cite{schleich:optics}. In the present paper we explore the transformation to the basis formed by the eigenstates of the coin operator further and apply it to the two families of three-state quantum walks introduced in \cite{stef:cont:def}.

The paper is organized as follows: In Section~\ref{sec2} we analyze the family of quantum walks constructed in \cite{stef:cont:def} as a parametrization of eigenvectors of the Grover coin. We show that the results \cite{falkner,machida} simplify considerably when the initial coin state is expressed in basis formed by eigenvectors of the coin operator. This allows us to determine the dependence of the moments of the distribution on the initial coin state. We also discuss the extremal regimes of quantum walks in consideration. In Section~\ref{sec3} we turn to the second family of quantum walks from \cite{stef:cont:def} which was constructed as a parametrization of eigenvalues of the Grover matrix. The details of the derivation of the localization probability and the group-velocity density are left for the Appendix~\ref{app:a}. Finally, we conclude and present an outlook in Section~\ref{sec4}.

\section{Eigenvector family}
\label{sec2}

Let us begin with a family of quantum walks which was introduced in \cite{stef:cont:def} and recently analyzed in \cite{falkner,machida}. The coin operators are in the standard basis of the coin space $\left\{|L\rangle,\ |S\rangle,\ |R\rangle\right\}$ given by the following matrix \cite{machida:note}
\begin{equation}
\label{coin:rho}
C(\rho) = \left(
  \begin{array}{ccc}
    -\rho^2 & \rho\sqrt{2-2\rho^2} & 1-\rho^2 \\
    \rho\sqrt{2-2\rho^2} & 2\rho^2-1 & \rho\sqrt{2-2\rho^2} \\
    1-\rho^2 & \rho\sqrt{2-2\rho^2} & -\rho^2 \\
  \end{array}
\right),
\end{equation}
with the coin parameter $\rho\in (0,1)$. We exclude the boundary points from our consideration since they result in trivial walks. Indeed, for $\rho=0$ the coin (\ref{coin:rho}) reduces to a permutation matrix with an additional phase shift to the state $|S\rangle$. Walk with such a coin can merely hop back and forth between the origin and its nearest neighbours. For the choice of $\rho=1$ the coin (\ref{coin:rho}) reduces to an identity matrix with an additional phase shift to the states $|L\rangle$ and $|R\rangle$. Such a coin is not mixing the coin states and the walk is simple - the $|L\rangle$ component of the initial coin state keeps hopping to the left, the $|R\rangle$ component keeps hopping to the right and the $|S\rangle$ component remains at the origin. Both quantum walks with $\rho=0$ and $\rho = 1$ can be analyzed in a straightforward way without the need for the week-limit theorems.

The set of coin operators (\ref{coin:rho}) was constructed in \cite{stef:cont:def} by a special parametrization of the eigenvectors of the 3x3 Grover matrix which is a member of this family corresponding to the choice of the coin parameter $\rho=\frac{1}{\sqrt{3}}$. One can show \cite{stef:cont:def} that the evolution operators of the quantum walks with the coin (\ref{coin:rho}) have a non-empty point spectrum for all values of $\rho\in(0,1)$. Consequently, the quantum walks show the localization effect in a similar way as found originally for the three-state Grover walk \cite{inui:pre,inui:psa}. The coin parameter $\rho$ determines directly the peak velocity of the walk \cite{stef:cont:def}, i.e. the positions of the peaks after $t$ steps of the quantum walk will be $\pm\rho t$.

The group-velocity density of quantum walks with the coin (\ref{coin:rho}) was recently derived by Falkner and Boettcher \cite{falkner} for the special case of the Grover walk corresponding to $\rho=\frac{1}{\sqrt{3}}$, and by Machida \cite{machida} for a general value of $\rho$. For a particle starting the walk from the origin with the initial coin state $|\psi_C\rangle$ the density has the form
\begin{equation}
\label{rho:machida}
w(v) = \frac{\sqrt{1-\rho^2}(d_0 + d_1 v + d_2 v^2)}{2\pi(1-v^2)\sqrt{\rho^2-v^2}}.
\end{equation}
With the group-velocity density (\ref{rho:machida}) one can determine the asymptotic value of the re-scaled moments of the particle's position $m$ in the limit of large number of steps $t$. Namely, the following relations hold for all $n\in\mathds{N}$
\begin{equation}
\label{rho:moment}
\lim\limits_{t\rightarrow +\infty} \left\langle \left(\frac{m}{t}\right)^n\right\rangle = \langle v^n\rangle = \int\limits_{-\rho}^\rho v^n\ w(v)\ dv.
\end{equation}
The group-velocity density can be also used to approximate the probability distribution after a finite number of steps $t$.  We have to replace $v$ in (\ref{rho:machida}) with $\frac{m}{t}$ and simultaneously re-normalize the distribution by $\frac{1}{t}$. This will be applied later in the Figures where we will compare the analytical results with a finite time numerical simulation.

\begin{figure}[h]
\includegraphics[width=0.4\textwidth]{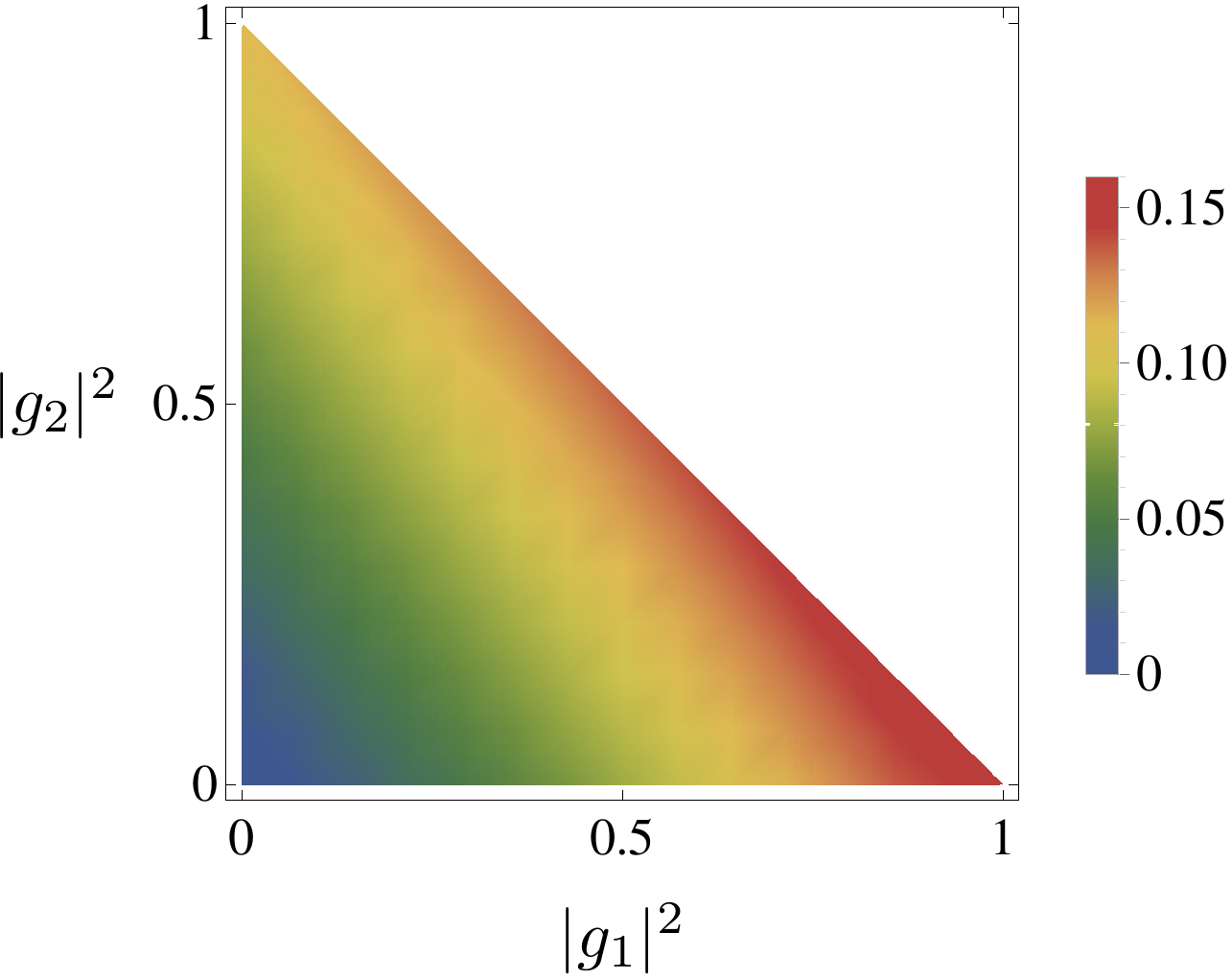}
\caption{(Color
  online) Second moment of the three-state walk with the coin operator (\ref{coin:rho}) as a function of $|g_1|^2$ and $|g_2|^2$. We have chosen the coin parameter $\rho=0.5$. The plot indicates that the greatest variance is achieved for the initial state $|\sigma_1^-\rangle$, while the smallest results from $|\sigma^+\rangle$. The domain of the plot is restricted to the lower triangle since the probabilities $|g_1|^2$ and $|g_2|^2$ are limited by the normalization condition (\ref{norm:init:rho}).}
\label{fig0}
\end{figure}

Note that the density (\ref{rho:machida}) looks relatively simple. However, we have yet to specify the terms $d_i$ which involve the dependence on the initial coin state $|\psi_C\rangle$. Machida \cite{machida} found that they are given by
\begin{eqnarray}
\label{d:machida}
\nonumber d_0 & = & |\alpha+\gamma|^2 + 2|\beta|^2,\\
\nonumber d_1 & = & 2\left\{\frac{}{}-|\alpha-\beta|^2+|\gamma-\beta|^2-\right.\\
& & \left. -\left(2-\frac{\sqrt{2-2\rho^2}}{\rho}\right){\rm Re}\left((\alpha-\gamma)\overline{\beta}\right) \right\},\\
\nonumber d_2 & = & |\alpha|^2 - 2|\beta|^2 + |\gamma|^2 -\\
\nonumber & & -2\left\{\frac{\sqrt{2-2\rho^2}}{\rho}{\rm Re}\left((\alpha+\gamma)\overline{\beta}\right) + \frac{2-\rho^2}{\rho^2}{\rm Re}\left(\alpha\overline{\gamma}\right)\right\}
\end{eqnarray}
where $\alpha, \beta$ and $\gamma$ are the coefficients of $|\psi_C\rangle$ in the standard basis, i.e.
\begin{equation}
\label{psi:stand}
|\psi_C\rangle = \alpha|L\rangle + \beta|S\rangle + \gamma|R\rangle.
\end{equation}
However, if we decompose the initial coin state in a more suitable basis it will simplify the relations (\ref{d:machida}) considerably. It can be anticipated that such a suitable basis is the one formed by the eigenstates of the coin operator. The eigenvectors of the coin operator (\ref{coin:rho}) read \cite{stef:cont:def}
\begin{eqnarray}
\label{rho:eigen}
\nonumber |\sigma^+\rangle & = & \sqrt{\frac{1 - \rho^2}{2}}|L\rangle +\rho|S\rangle + \sqrt{\frac{1 - \rho^2}{2}}|R\rangle, \\
\nonumber |\sigma_1^-\rangle & = & \frac{\rho}{\sqrt{2}}|L\rangle -\sqrt{1-\rho^2}|S\rangle + \frac{\rho}{\sqrt{2}}|R\rangle, \\
|\sigma_2^-\rangle & = & \frac{1}{\sqrt{2}}(|L\rangle - |R\rangle).
\end{eqnarray}
They satisfy the eigenvalue equations
$$
C(\rho)|\sigma^+\rangle = |\sigma^+\rangle,\qquad C(\rho)|\sigma_i^-\rangle = -|\sigma_i^-\rangle, i=1,2.
$$
We decompose the initial coin state into the eigenstate basis in the form
\begin{equation}
\label{psi:c:rho}
|\psi_C\rangle = g_+|\sigma^+\rangle + g_1|\sigma_1^-\rangle + g_2|\sigma_2^-\rangle,
\end{equation}
where the probability amplitudes $g_+$ and $g_{1,2}$ are restricted by the normalization condition
\begin{equation}
\label{norm:init:rho}
|g_+|^2 + |g_1|^2 + |g_2|^2 = 1.
\end{equation}
From the relations (\ref{rho:eigen}) we find that the coefficients in the standard basis $\alpha,\beta,\gamma$ are in the eigenstate basis given by
\begin{eqnarray}
\nonumber \alpha & = & \frac{1}{\sqrt{2}}\left(\sqrt{1-\rho^2} g_+ + \rho g_1 + g_2\right),\\
\nonumber \beta & = & g_+ \rho - \sqrt{1 - \rho^2}g_1,\\
\nonumber \gamma & = & \frac{1}{\sqrt{2}}\left(\sqrt{1-\rho^2} g_+ + \rho g_1 - g_2\right).
\end{eqnarray}
Plugging these expressions into (\ref{d:machida}) and using the normalization condition (\ref{norm:init:rho}) we obtain much more convenient formulas for the terms $d_i$, namely
\begin{eqnarray}
\nonumber d_0 & = & 2(1-|g_2|^2),\\
\nonumber d_1 & = & -\frac{2}{\rho}\left(g_1\overline{g_2} + \overline{g_1}g_2\right),\\
d_2 & = & \frac{2}{\rho^2}\left(|g_1|^2 + 2|g_2|^2 - 1\right).
\end{eqnarray}
Finally, the group-velocity density reads
\begin{eqnarray}
\label{dist:rho}
w(v) & = & \frac{\sqrt{1-\rho^2}}{\pi(1-v^2)\sqrt{\rho^2-v^2}} \left(1-|g_2|^2\frac{}{} - \right.\\
\nonumber & & \left. - (g_1\overline{g_2} + \overline{g_1}g_2)\frac{v}{\rho} + (|g_1|^2 + 2|g_2|^2 - 1)\frac{v^2}{\rho^2}\right).
\end{eqnarray}
This result allows us to determine the dependence of the moments (\ref{rho:moment}) on the initial coin state in a straightforward way. Namely, odd moments of the group-velocity have the form
$$
\langle v^{2n+1}\rangle =  O_n(\rho)\left(g_1\overline{g_2} + \overline{g_1}g_2\right),
$$
where we have denoted
$$
O_n(\rho) = -\frac{\sqrt{1-\rho^2}}{\rho}\int\limits_{-\rho}^\rho \frac{v^{2n+2}}{\pi(1-v^2)\sqrt{\rho^2-v^2}}\ dv.
$$
We see that in the eigenstate basis the odd moments are determined by the coherent combination of the probability amplitudes $g_1$ and $g_2$. On the other hand, the even moments (\ref{rho:moment}) depend only on the probabilities $|g_1|^2$ and $|g_2|^2$ of finding the particle initially in the coin state $|\sigma_1^-\rangle$ or $|\sigma_2^-\rangle$. This means that the mixed initial coin state of the form
$$
\rho_C = |g_+|^2|\sigma^+\rangle\langle\sigma^+| + |g_1|^2|\sigma_1^-\rangle\langle\sigma_1^-| + |g_2|^2|\sigma_2^-\rangle\langle\sigma_2^-|,
$$
results in a distribution with the same even moments as the pure coin state (\ref{psi:c:rho}). In particular, the second moment is given by
\begin{equation}
\label{v2:rho}
\langle v^2\rangle = \left(\left|g_1\right|^2+1\right)\Delta_1(\rho) + \left(\left|g_2\right|^2-1\right)\Delta_2(\rho),
\end{equation}
where we have used the notation
\begin{eqnarray}
\nonumber \Delta_1(\rho) & = & \frac{1+\rho^2-\sqrt{1-\rho^2}}{2 + 2\sqrt{1-\rho^2}},\\
\nonumber \Delta_2(\rho) & = & \frac{2-\rho^2-2 \sqrt{1-\rho^2}}{\rho^2}.
\end{eqnarray}
One can easily check that the inequalities
$$
\Delta_1(\rho) > \Delta_2(\rho) > 0,
$$
hold for all $\rho\in(0,1)$. It is then straightforward to show that the state giving rise to the distribution with the smallest variance is the one corresponding to $g_1=g_2=0$, i.e. the eigenstate $|\sigma^+\rangle$. Analogously, the eigenstate $|\sigma_1^-\rangle$ yields the distribution with the greatest variance. We display the second moment (\ref{v2:rho}) as a function of the probabilities $|g_1|^2$ and $|g_2|^2$ in Fig.~\ref{fig0} for the choice of the coin parameter $\rho=0.5$. The plot indicates that $|\sigma^+\rangle$ yields the smallest variance while $|\sigma_1^-\rangle$ gives the greatest.


\begin{figure}[htbp]
\begin{center}
\includegraphics[width=0.45\textwidth]{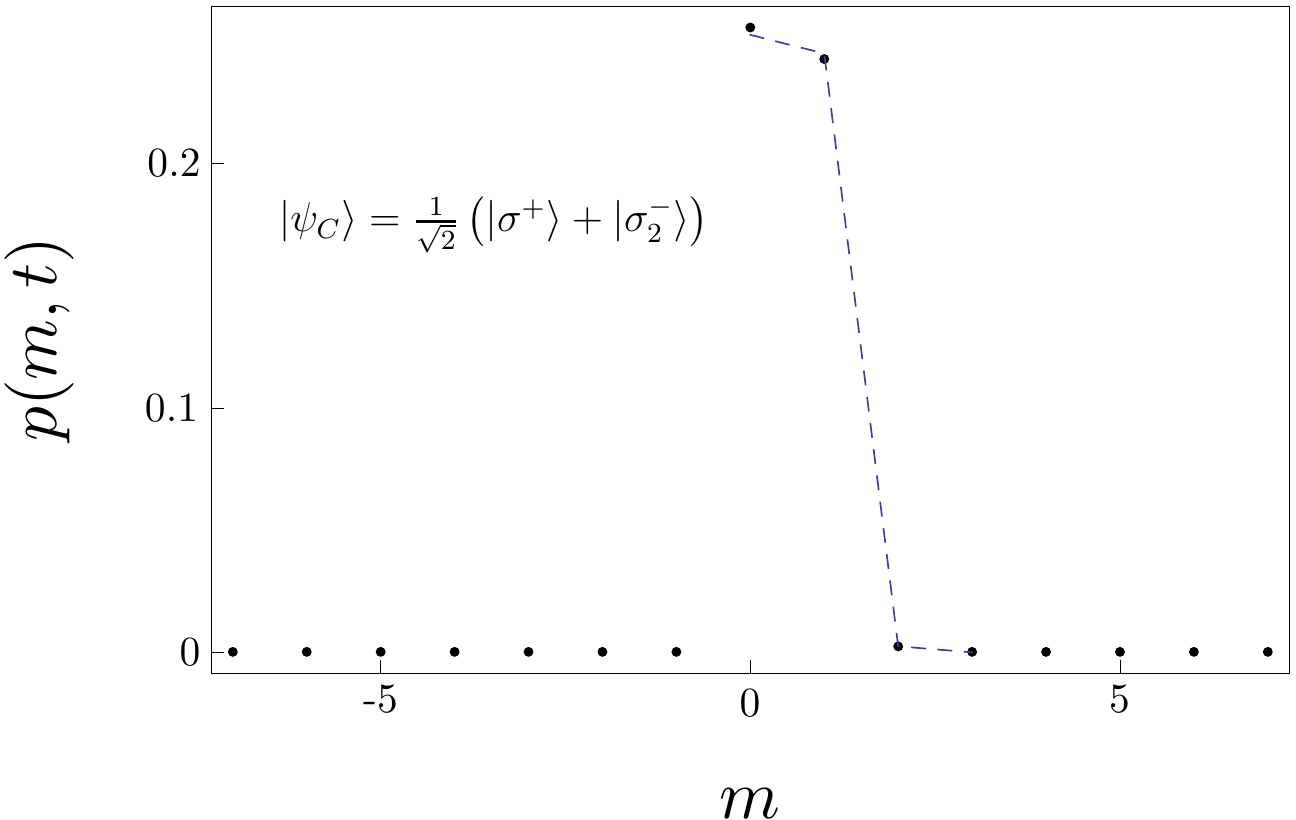}
\end{center}
\caption{(Color
  online) Probability distribution in the vicinity of the origin for the three-state Grover walk starting with the initial state $|\psi_C\rangle = \frac{1}{\sqrt{2}}\left(|\sigma^+\rangle + |\sigma_2^-\rangle\right)$ after $t=1000$ steps. The localization depicted by the blue-dashed line appears only for positive $m$, as predicted by (\ref{loc:pos}).}
\label{fig:loc}
\end{figure}

We note that the group-velocity density (\ref{dist:rho}) is not normalized to unity, as was found in \cite{falkner,machida}. Indeed, we obtain the following result
\begin{equation}
\label{rho:total:dens}
\int\limits_{-\rho}^\rho w(v)dv = 1-|g_2|^2 - \frac{\sqrt{1-\rho^2}-1}{\rho^2} \left(|g_1|^2 + 2|g_2|^2 - 1\right).
\end{equation}
The remaining part of the probability is in the exponential peak corresponding to localization. This was also recently calculated by Falkner and Boettcher \cite{falkner} for $\rho=\frac{1}{\sqrt{3}}$ corresponding to the Grover walk, and by Machida \cite{machida} for a general value of $\rho$. The probability to find the particle at position $m$ in the limit $t\rightarrow +\infty$ is given by \cite{machida}
\begin{eqnarray}
\nonumber p_\infty(m) & = & \frac{1}{128(1-\rho^2)^2}\left\{2(1-\rho^2)\left|B\nu^{|m+1|} + A\nu^{|m|}\right|^2 + \right.\\
\nonumber  & & \left. + \rho^2\left|B\nu^{|m+1|} + (A+B)\nu^{|m|} + A\nu^{|m-1|} \right|^2 + \right. \\
\nonumber & & \left. 2(1-\rho^2)\left|B\nu^{|m|} + A\nu^{|m-1|}\right|^2 \right\}.
\end{eqnarray}
Here $\nu$ depends on the coin parameter
$$
\nu = -\frac{2-\rho^2-2\sqrt{1-\rho^2}}{\rho^2},
$$
and $A$, $B$ involves the initial coin state
\begin{eqnarray}
\nonumber A & = & 4\left(1-\rho^2\right) \alpha  + 2\rho \sqrt{2-2 \rho^2} \beta,\\
\nonumber B & = & 4\left(1-\rho^2\right) \gamma  + 2\rho \sqrt{2-2 \rho^2} \beta.
\end{eqnarray}
The simplification of this result by turning into the eigenvector basis (\ref{rho:eigen}) is perhaps even more significant than the one achieved for the group-velocity density (\ref{dist:rho}). Indeed, after some algebra we find that the localization probability $p_\infty(m)$ is given by
\begin{equation}
\label{loc:rho}
p_\infty(m) = \left\{
                \begin{array}{c}
                  \frac{2-2\rho^2}{\rho^4}\nu^{2m} |g_+ + g_2|^2 ,\quad m>0 , \\
                   \\
                   \frac{1}{\rho^2}|\nu|\left\{|g_+|^2 + (1-\rho^2)|g_2|^2 \right\}, \quad m = 0,\\
                   \\
                  \frac{2-2\rho^2}{\rho^4}\nu^{2|m|} |g_+ - g_2|^2 ,\quad m<0 \\
                \end{array}
              \right.
\end{equation}
This result shows that the central peak is indeed an exponential with the base $\nu^2$, except for the origin. Moreover, the dependence on the initial coin state is particularly simple in the eigenvector basis. For the origin the localization probability is given by an incoherent combination of $|g_+|^2$ and $|g_2|^2$, while outside the dependence is determined by a coherent combination of amplitudes $g_+$ and $g_2$. Moreover, the dependence differs for positive and negative $m$. This fact can be exploited to force the particle to localize only on the positive or negative half-line by a proper choice of the initial coin state. As an example, consider the initial coin state
$$
|\psi_C\rangle = \frac{1}{\sqrt{2}}\left(|\sigma^+\rangle + |\sigma_2^-\rangle\right).
$$
Using the expression (\ref{loc:rho}) we find that in this case the localization probability equals
\begin{equation}
\label{loc:pos}
p_\infty(m) = \left\{\begin{array}{c}
                \frac{4-4\rho^2}{\rho^4}\nu^{2m},\quad m>0 \\
                \ \\
                \left(\frac{1}{\rho^2}- \frac{1}{2}\right)|\nu|,\quad m=0 \\
                \ \\
               0,\quad m<0
              \end{array}\right.
\end{equation}
Localization thus appears only on the positive half-line. We illustrate this effect in Figure~\ref{fig:loc} where we present the probability distribution of the Grover walk corresponding to $\rho=\frac{1}{\sqrt{3}}$ after $t=1000$ steps. To unravel the unusual behaviour of localization we display only a small neighbourhood of the origin.


We note that one can easily check that
$$
\sum\limits_{m=-\infty}^\infty p_\infty(m) = |g_2|^2 + \frac{\sqrt{1-\rho^2}-1}{\rho^2} \left(|g_2|^2-|g_+|^2\right).
$$
This together with (\ref{rho:total:dens}) and the normalization condition (\ref{norm:init:rho}) results in
$$
\sum\limits_{m=-\infty}^\infty p_\infty(m) + \int\limits_{-\rho}^\rho w(v)\ dv = 1,
$$
i.e. the complete probability density is properly normalized to unity.

Let us now illustrate our findings on several examples. A typical distribution resulting from the three-state walk with the coin (\ref{coin:rho}) has three characteristic peaks. Two are on the edges of the distribution and correspond to the divergency of the group-velocity density (\ref{dist:rho}) for $v$ approaching $\pm\rho$. The third one is the exponential peak at the origin corresponding to localization (\ref{loc:rho}). However, these characteristics will be altered considerably when we chose one of the eigenvectors of the coin operator as the initial state of the walk.

\begin{figure}[h]
\includegraphics[width=0.45\textwidth]{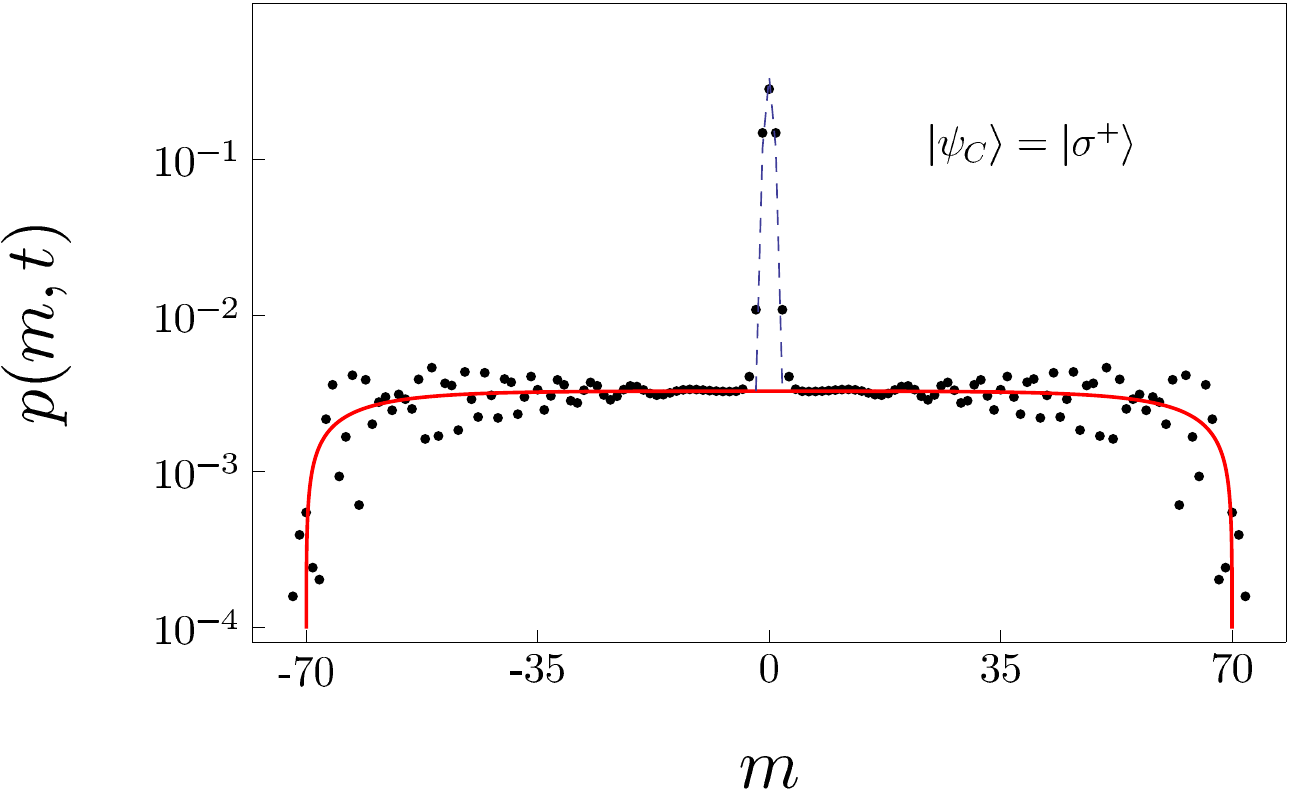}
\caption{(Color
  online) Probability distribution of the three-state walk with the coin parameter $\rho=0.7$  after $t=100$ steps. As the initial state we have chosen the coin eigenstate $|\sigma^+\rangle$. For this initial coin state both peaks on the edges of the distribution vanish. This corresponds to the fact that the group-velocity density (\ref{dist:rho:hp}), depicted by the red curve, tends to zero for $v$ approaching $\pm\rho$. To unravel this feature we plot the probability distribution on the logarithmic scale. The blue-dashed line depicts the localization probability (\ref{loc:rho}).}
\label{fig1}
\end{figure}

First, consider the eigenstate $|\sigma^+\rangle$ as the initial coin state of the walk. From the relation (\ref{dist:rho}) we find that the group-velocity density is given by
\begin{equation}
\label{dist:rho:hp}
w_{|\sigma^+\rangle}(v) = \frac{\sqrt{1-\rho^2}\sqrt{\rho^2-v^2}}{\pi\rho^2(1-v^2)}.
\end{equation}
We see that the density does not diverge for $v$ approaching $\pm\rho$. Hence, both peaks on the edges of the distribution disappear. We illustrate this effect in Fig.~\ref{fig1}, where we display the distribution after 100 steps. The coin parameter was chosen as $\rho=0.7$.


\begin{figure}[h]
\includegraphics[width=0.45\textwidth]{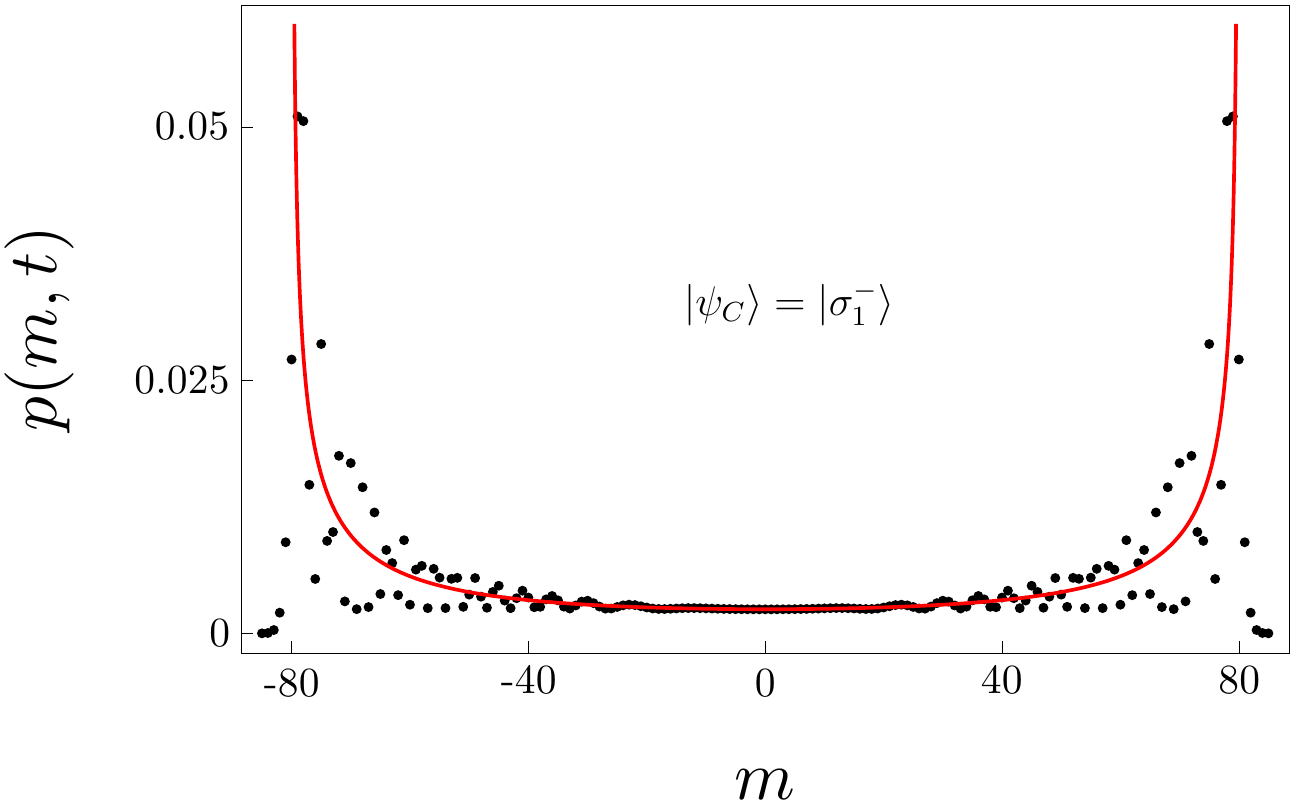}
\caption{(Color
  online) Probability distribution of the three-state walk with the coin parameter $\rho=0.8$  after $t=100$ steps. As the initial state we have chosen the coin eigenstate $|\sigma_1^-\rangle$. For this initial coin state the localization effect disappears. The red curve corresponds to the density (\ref{dist:rho}).}
\label{fig2}
\end{figure}

Let us turn to the eigenstate $|\sigma_1^-\rangle$. From the relation (\ref{loc:rho}) we see that for the choice of the parameters $g_1=1$ and $g_2 = g_+ = 0$  the localization probability equals zero for all $m$. Hence, for this initial state the localization disappears, in accordance with the findings of \cite{machida}. We illustrate this effect in Fig.~\ref{fig2}, where we show the distribution after 100 steps and the coin parameter $\rho=0.8$. We can clearly see that there are only two peaks on the edges of the distribution. Consequently, the state $|\sigma_1^-\rangle$ yields the distribution with the greatest variance, as indicated by Fig.~\ref{fig0}.


\begin{figure}
\includegraphics[width=0.45\textwidth]{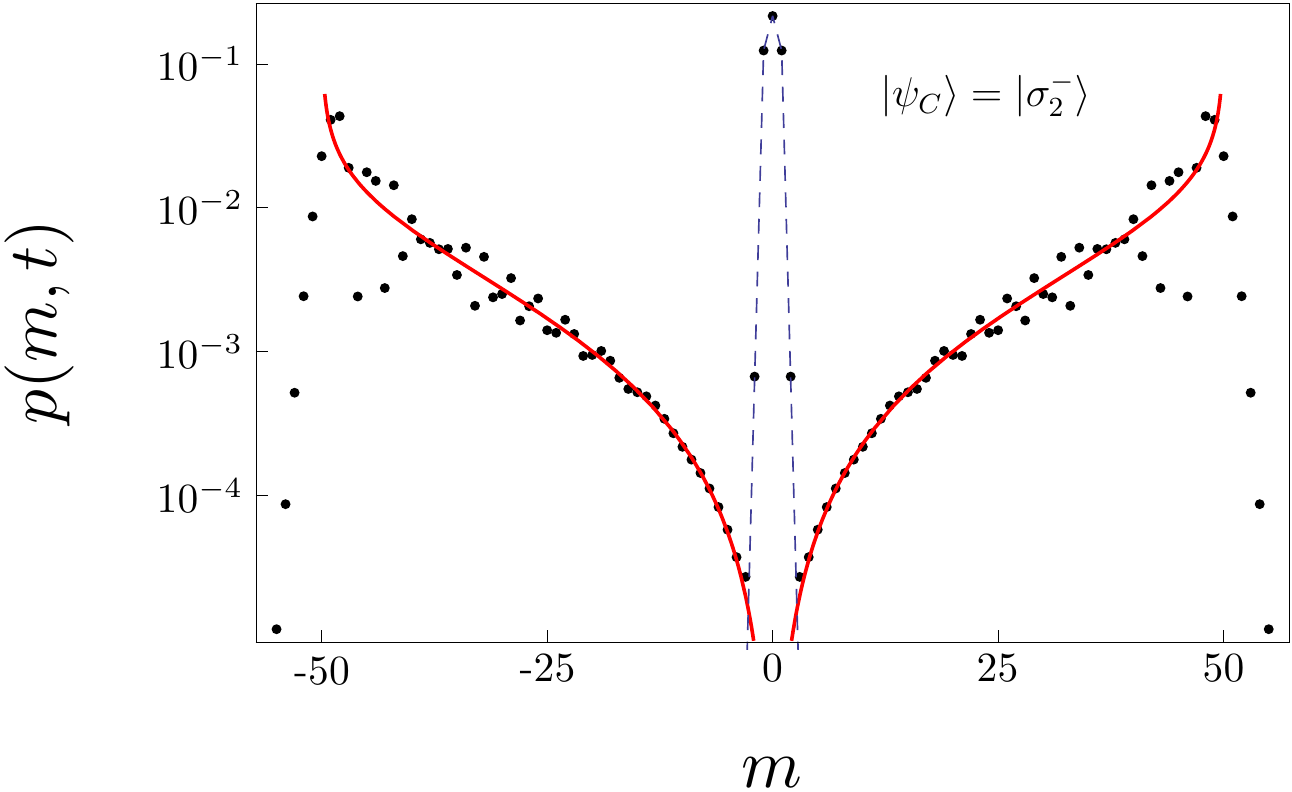}
\caption{(Color
  online) Probability distribution of the three-state walk with the coin parameter $\rho=0.5$  after $t=100$ steps. As the initial state we have chosen the coin eigenstate $|\sigma_2^-\rangle$. The probability density (\ref{dist:rho:h2}) depicted by the red curve tends to zero near the origin. To unravel this effect we use logarithmic scale on the $y$-axis. The blue-dashed line corresponds to the localization probability (\ref{loc:rho}).}
\label{fig3}
\end{figure}

Concerning the last eigenstate $|\sigma_2^-\rangle$, we find that the group-velocity density is given by
\begin{equation}
\label{dist:rho:h2}
w_{|\sigma_2^-\rangle}(v) = \frac{\sqrt{1-\rho^2}v^2}{\pi\rho^2(1-v^2)\sqrt{\rho^2-v^2}}.
\end{equation}
We see that the density vanishes as $v$ tends to zero. To illustrate this effect we plot in Fig.~\ref{fig3} the distribution after 100 steps on a logarithmic scale. The coin parameter was chosen as $\rho=0.5$. The plot indicates that the distribution tends to zero around the origin, except for a very small neighbourhood where the localization dominates.


\begin{figure}[h]
\includegraphics[width=0.45\textwidth]{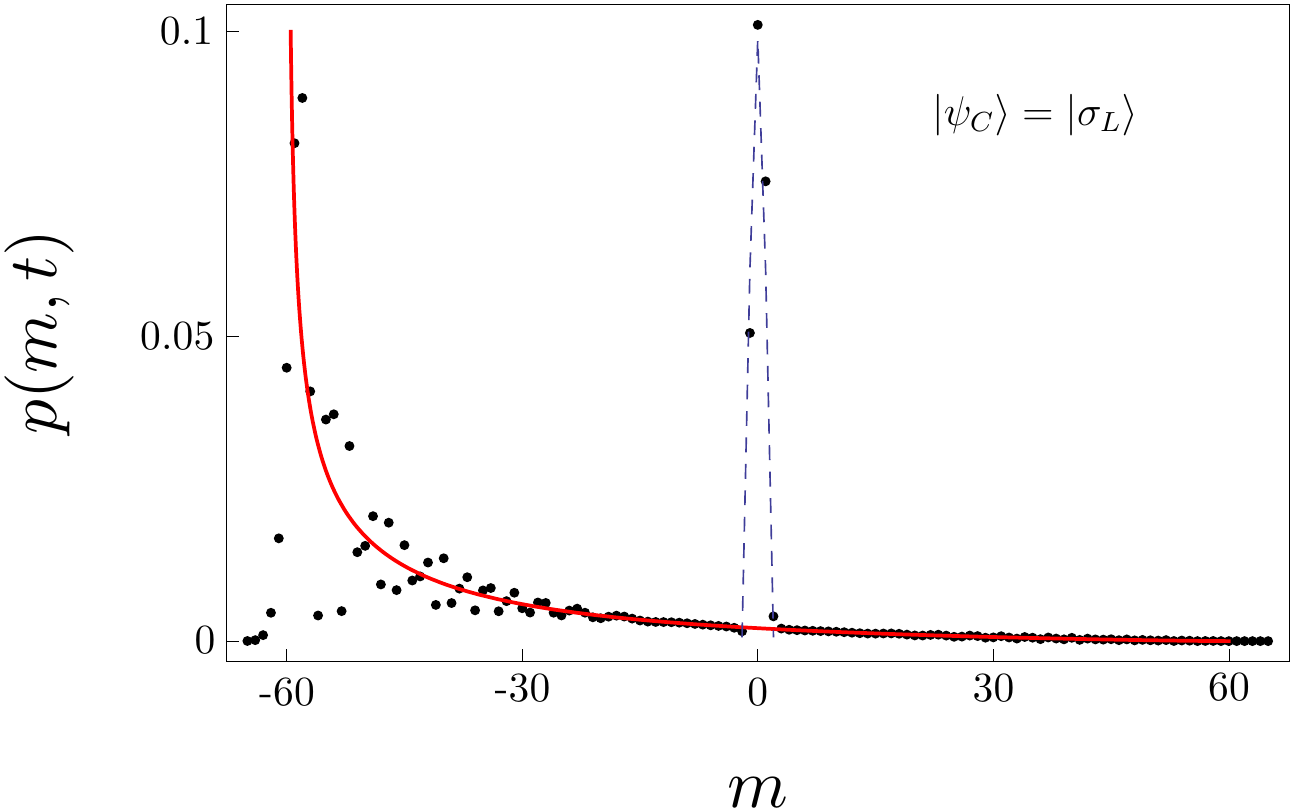}
\caption{(Color
  online) Probability distribution of the three-state walk with the coin parameter $\rho=0.6$  after $t=100$ steps. As the initial coin state we have chosen $|\sigma_L\rangle$ given by equation (\ref{sigma:L}). In this case the peak on the right hand side of the lattice disappears, as predicted by the density (\ref{dist:sigma:L}) illustrated with the red curve. The blue-dashed line depicts the localization probability (\ref{loc:rho}).}
\label{fig4}
\end{figure}

We have seen that for the initial coin state $|\sigma^+\rangle$ both peaks on the edges of the distribution vanish. However, it is possible to construct a state for which only one of the peaks disappears. Indeed, consider the coin state
\begin{equation}
\label{sigma:L}
|\sigma_L\rangle = \frac{1}{\sqrt{2}}\left(|\sigma_1^-\rangle + |\sigma_2^-\rangle\right).
\end{equation}
From the relation (\ref{dist:rho}) we find that the group-velocity density for this initial state reads
\begin{equation}
\label{dist:sigma:L}
w_{|\sigma_L\rangle}(v) = \frac{\sqrt{1-\rho^2}\sqrt{(\rho-v)^3}}{2\pi\rho^2(1-v^2)\sqrt{\rho+v}}.
\end{equation}
Such a density tends to zero for $v$ approaching $\rho$. Nevertheless, the divergency at $v=-\rho$ remains. We illustrate this feature in Fig.~\ref{fig4} where we display the distribution after 100 steps for the coin parameter $\rho=0.6$.


To conclude this Section, we note that the characteristic features of this family of quantum walks are maintained for all values of the coin parameter $\rho\in(0,1)$, which scales the rate at which the walk spreads through the lattice. The three-state Grover walk is a particular case corresponding to the case $\rho = \frac{1}{\sqrt{3}}$. The change of the basis of the coin space to the one formed by the eigenstates of the coin operator (\ref{coin:rho}) allowed us to simplify the description of the walk considerably. Consequently, interesting dynamical regimes which are otherwise hidden were easily identified.


\section{Eigenvalue family}
\label{sec3}

Let us turn to the second family of quantum walks we have introduced in \cite{stef:cont:def}. The coin operators are in the standard basis of the coin space represented by the following matrix
\begin{equation}
\label{coin:phi}
C(\varphi) =
\frac{1}{6}\left(
\begin{array}{ccc}
 -1-e^{2i\varphi} & 2(1+e^{2i\varphi}) & 5-e^{2i\varphi} \\
  2(1+e^{2i\varphi}) & 2(1-2e^{2i\varphi}) & 2(1+e^{2i\varphi}) \\
   5-e^{2i\varphi} & 2(1+e^{2i\varphi}) & -1-e^{2i\varphi} \\
   \end{array}
   \right),
\end{equation}
with the coin parameter $\varphi\in\langle 0,\frac{\pi}{2})$. For the choice of $\varphi=0$ the coin operator turns to the 3x3 Grover matrix. We exclude the other boundary point $\varphi=\frac{\pi}{2}$ from our consideration since in such a case the coin operator (\ref{coin:phi}) reduces to a permutation matrix and the resulting quantum walk is trivial.

The set of matrices (\ref{coin:phi}) was constructed in \cite{stef:cont:def} as a special parametrization of the eigenvalues of the 3x3 Grover matrix. It was shown that the evolution operators of quantum walks with the coin operators (\ref{coin:phi}) maintain a point spectrum for all values of $\varphi\in\langle 0,\frac{\pi}{2})$. Hence, the family of quantum walks driven by such coins show the localization effect. The parameter $\varphi$ determines the rate at which the walk spreads through the lattice. Namely, we have shown in \cite{stef:cont:def} that the positions of the peaks after $t$ steps of the walk will be $\pm\eta t$, where the peak velocity $\eta$ is given by
\begin{equation}
\label{phi:peak:vel}
\eta = \frac{1}{\sqrt{6}}\sqrt{3-\cos^2{\varphi}-\sin{\varphi}\sqrt{9-\cos^2{\varphi}}}.
\end{equation}

Before we proceed with the analysis of this family of quantum walks, we first choose a suitable basis in the coin space. We again employ the eigenstates of the coin operator (\ref{coin:phi}) which are given by
\begin{eqnarray}
\nonumber |\gamma^+\rangle & = & \frac{1}{\sqrt{3}}\left(|L\rangle + |S\rangle + |R\rangle\right),\\
\nonumber |\gamma_1^-\rangle & = & \frac{1}{\sqrt{6}}\left(|L\rangle - 2|S\rangle + |R\rangle \right),\\
\nonumber |\gamma_2^-\rangle & = & \frac{1}{\sqrt{2}}\left(|L\rangle - |R\rangle\right).
\end{eqnarray}
They satisfy the following eigenvalue equations
\begin{eqnarray}
\nonumber C(\varphi)|\gamma^+\rangle & = & |\gamma^+\rangle,\\
\nonumber C(\varphi)|\gamma_{1}^-\rangle & = & -e^{2 i\varphi}|\gamma_{1}^-\rangle,\\
\nonumber C(\varphi)|\gamma_{2}^-\rangle & = & -|\gamma_{2}^-\rangle.
\end{eqnarray}
We decompose the initial coin state $|\psi_C\rangle$ into the eigenvector basis in the following form
\begin{equation}
\label{init:phi}
|\psi_C\rangle = g_+|\gamma^+\rangle + g_1|\gamma_1^-\rangle + g_2|\gamma_2^-\rangle,
\end{equation}
where the probability amplitudes $g_+$ and $g_{1,2}$ satisfy the normalization condition (\ref{norm:init:rho}).

\begin{figure}[h]
\includegraphics[width=0.4\textwidth]{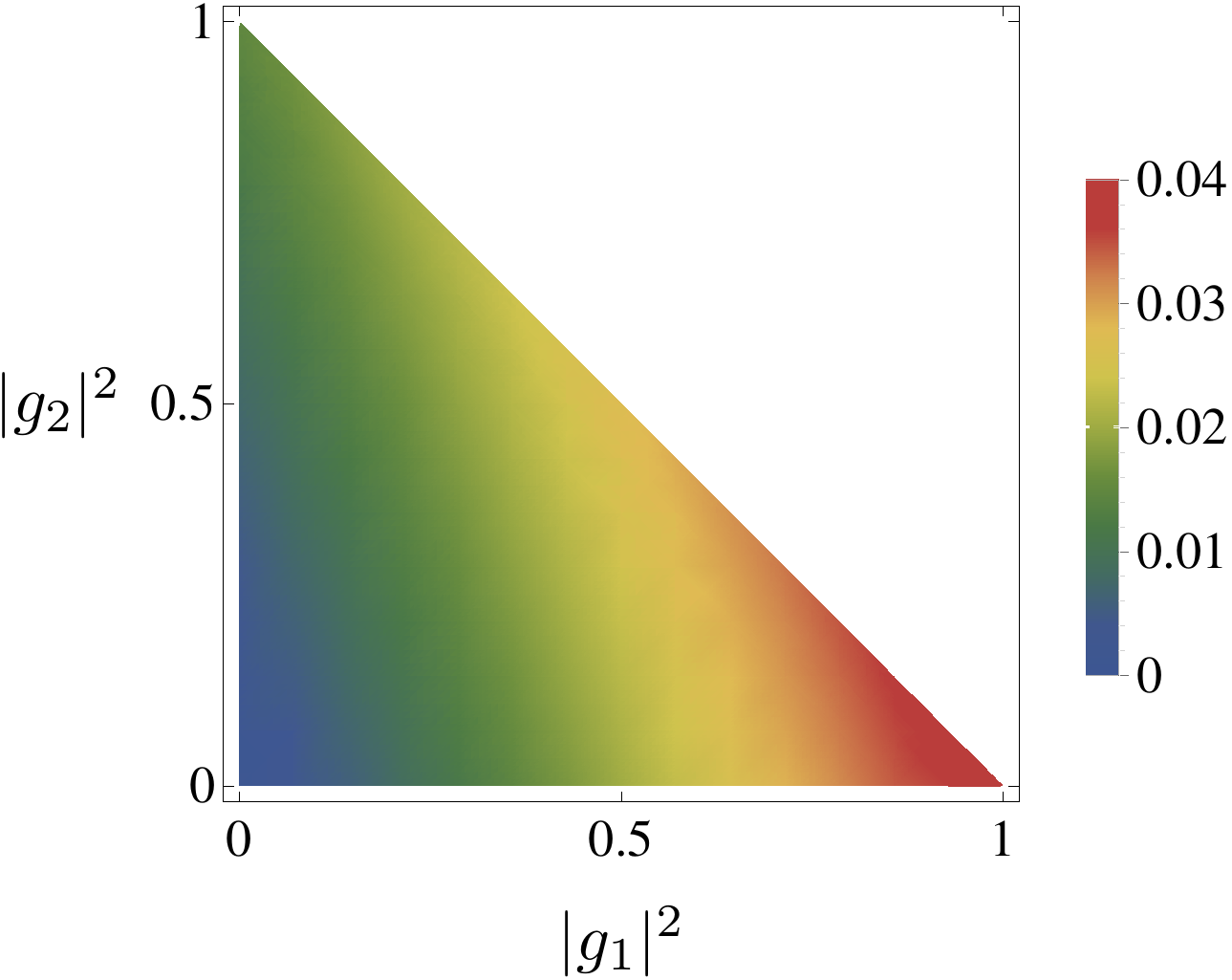}
\caption{(Color
  online) Second moment of the three-state walk with the coin (\ref{coin:phi}) as a function of $|g_1|^2$ and $|g_2|^2$. We have chosen the coin parameter $\varphi=\frac{\pi}{4}$. Note the different scale in comparison with Fig.~\ref{fig0}. The domain of the plot is restricted due to the normalization condition (\ref{norm:init:rho}).}
\label{fig:phi:1}
\end{figure}

Let us now turn to the group-velocity density of the family of quantum walks with coin operators (\ref{coin:phi}). We leave the details of the derivation for the Appendix, where we show that it can be expressed in the following form
\begin{eqnarray}
\label{dist:phi}
\nonumber w(v) & = & \frac{1}{6\pi (1-v^2) \Theta}\left[\frac{}{}\left(3|g_1|^2 + 5|g_2|^2 - 2\right)\Lambda_+ +\right. \\
\nonumber & &\left. + \left(1 - |g_1|^2 - 2|g_2|^2\right) \Omega - \right.\\
\nonumber & & \left. -  \sqrt{3}v(g_1\overline{g_2} + \overline{g_1}g_2 + i(g_1\overline{g_2} -\overline{g_1}g_2)\tan\varphi)\Lambda_+ + \right.\\
 & &\left. + i v(g_2\overline{g_+}-\overline{g_2}g_+) \Xi\frac{}{}\right].
\end{eqnarray}
Here we have denoted
\begin{eqnarray}
\label{phi:terms}
\nonumber \Lambda_\pm & = & \Phi_+ \pm \Phi_-,\\
\nonumber \Phi_\pm & = & \sqrt{9(1-v^2) - (5+3v^2)\cos^2\varphi \pm 12 \Theta \cos\varphi },\\
\nonumber \Omega & = & 4\cos\varphi \frac{(5-3v^2)\cos\varphi \Lambda_+ + 3\Theta \Lambda_-}{8\cos^2\varphi + 3v^2\sin^2\varphi},\\
\nonumber \Xi & = & 3\sqrt{6}\tan\varphi \frac{(v^2 + \cos^2\varphi)\Lambda_+ -  \Theta\cos\varphi \Lambda_-}{8\cos^2\varphi + 3v^2\sin^2\varphi}, \\
\nonumber \Theta & = & \sqrt{(\eta^2-v^2)\left(\eta^2-v^2+\sin\varphi\sqrt{1-\frac{\cos^2\varphi}{9}}\right)}.\\
\end{eqnarray}
The formula for the group-velocity (\ref{dist:phi}) is much more involved than the one derived in the previous Section  (see equation (\ref{dist:rho})). Nevertheless, the dependence on the initial coin state is still relatively simple in the eigenvector basis. In particular, one can show that the even moments depend only on the probabilities $|g_1|^2$ and $|g_2|^2$ of finding the particle initially in the coin state $|\gamma_1^-\rangle$ or $|\gamma_2^-\rangle$. Indeed, note that all expressions defined in (\ref{phi:terms}) are even functions of $v$. Consequently, the first two terms in the group-velocity density (\ref{dist:phi}) are even in $v$, while the remaining two are odd. Hence, only the first two terms of (\ref{dist:phi}) contribute to the calculation of even moments. As an example, for the second moment we obtain the result
\begin{eqnarray}
\label{meanv2:phi}
\nonumber \langle v^2\rangle & = & \left(3|g_1|^2 + 5|g_2|^2 - 2\right)\Delta_1(\varphi) +\\
& & + \left(1 - |g_1|^2-2|g_2|^2\right)\Delta_2(\varphi),
\end{eqnarray}
where $\Delta_i(\varphi)$ denotes the following integrals
\begin{eqnarray}
\nonumber \Delta_1(\varphi) & = & \int\limits_{-\eta}^\eta \frac{v^2}{6\pi (1-v^2) \Theta}\ \Lambda_+ \ dv ,\\
\nonumber \Delta_2(\varphi) & = & \int\limits_{-\eta}^\eta \frac{v^2}{6\pi (1-v^2) \Theta}\ \Omega \ dv,
\end{eqnarray}
which have to be evaluated numerically for a given value of $\varphi$. We display the second moment as a function of the probabilities $|g_1|^2$ and $|g_2|^2$ in Fig.~\ref{fig:phi:1}. The coin parameter $\varphi$ was chosen as $\frac{\pi}{4}$. The plot indicates that the greatest second moment is achieved for the eigenstate $|\gamma_1^-\rangle$, while the smallest results from $|\gamma_+\rangle$.


We note that the odd moments are determined by the last two terms of the group-velocity density (\ref{dist:phi}). Hence, they are influenced by the coherence of the initial coin state.

Similarly to the previous model the group-velocity density (\ref{dist:phi}) is not normalized since we find
\begin{equation}
\label{dist:phi:norm}
\int\limits_{-\eta}^\eta w(v)\ dv =\sqrt{6} - 2 + (3-\sqrt{6})|g_1|^2 + (5-2\sqrt{6})|g_2|^2,
\end{equation}
which differs from one unless $g_1=1$. The remaining part of the probability is in the localization. Concerning this part of the probability distribution we show in the Appendix that it is completely independent of $\varphi$. Hence, it coincides with the result for the three-state Grover walk  corresponding to the choice of $\varphi=0$. As we have mentioned earlier, the Grover walk also belongs to the eigenvector family we have studied in the previous Section. Using the result (\ref{loc:rho}) for $\rho=\frac{1}{\sqrt{3}}$ corresponding to the Grover walk we find that the localization probability for three-state quantum walks with the coin (\ref{coin:phi}) reads
\begin{equation}
\label{loc:phi}
p_\infty(m) = \left\{
                \begin{array}{c}
                  12(5-2\sqrt{6})^{2m}|g_+ + g_2|^2,\quad m>0, \\
                   \\
                  (5-2\sqrt{6})(3|g_+|^2 + 2|g_2|^2),\quad m = 0, \\
                   \\
                  12(5-2\sqrt{6})^{2|m|}|g_+ - g_2|^2,\quad m<0. \\
                \end{array}
              \right.
\end{equation}
We can easily check that
$$
\sum\limits_{m=-\infty}^\infty p_\infty(m) = (\sqrt{6}-2)|g_2|^2 + (3-\sqrt{6})|g_+|^2,
$$
which together with (\ref{dist:phi:norm}) and the normalization condition (\ref{norm:init:rho}) guarantees that
$$
\sum\limits_{m=-\infty}^\infty p_\infty(m) + \int\limits_{-\eta}^\eta w(v)\ dv = 1.
$$

Let us now illustrate our findings on several examples. For the choice of the coin parameter $\varphi=0$ the walk reduces to the three-state Grover walk and the group velocity density (\ref{dist:phi}) simplifies enormously to
\begin{eqnarray}
\label{dist:phi:grover}
\nonumber w(v) & = & \frac{\sqrt{2}}{\pi(1-v^2)\sqrt{1-3v^2}} \left(1-|g_2|^2\frac{}{} - \right.\\
\nonumber & & \left. - \sqrt{3}(g_1\overline{g_2} + \overline{g_1}g_2)v + 3(|g_1|^2 + 2|g_2|^2 - 1)v^2\right).
\end{eqnarray}
Note that this formula coincides with the result (\ref{dist:rho}) of the previous Section for the particular choice of $\rho=\frac{1}{\sqrt{3}}$. Hence, all features we have discussed in Section~\ref{sec2} apply to the choice of $\varphi=0$. In particular, the coin eigenstates will play a special role since they result in an extremal regime of the walk. For $\varphi>0$ the role of coin eigenstates will be less prominent and most of the features we have found in Section~\ref{sec2} diminish. The only exception is the behaviour of localization (\ref{loc:phi}) which is completely independent of $\varphi$. As a consequence, choosing the eigenvector $|\gamma_1^-\rangle$ as the initial coin state of the walk will cancel localization for any value of the coin parameter $\varphi$.

\begin{figure}[h]
\includegraphics[width=0.45\textwidth]{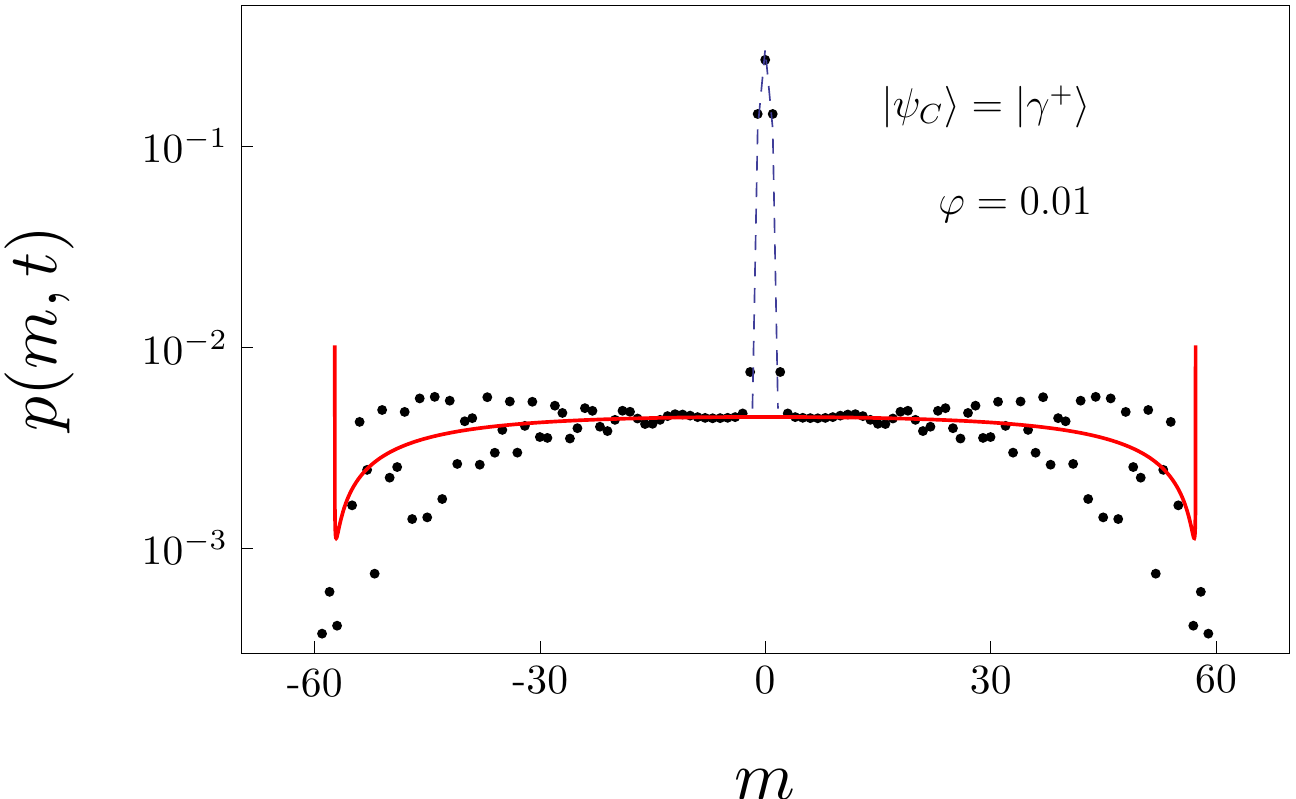}\vspace{12pt}
\includegraphics[width=0.45\textwidth]{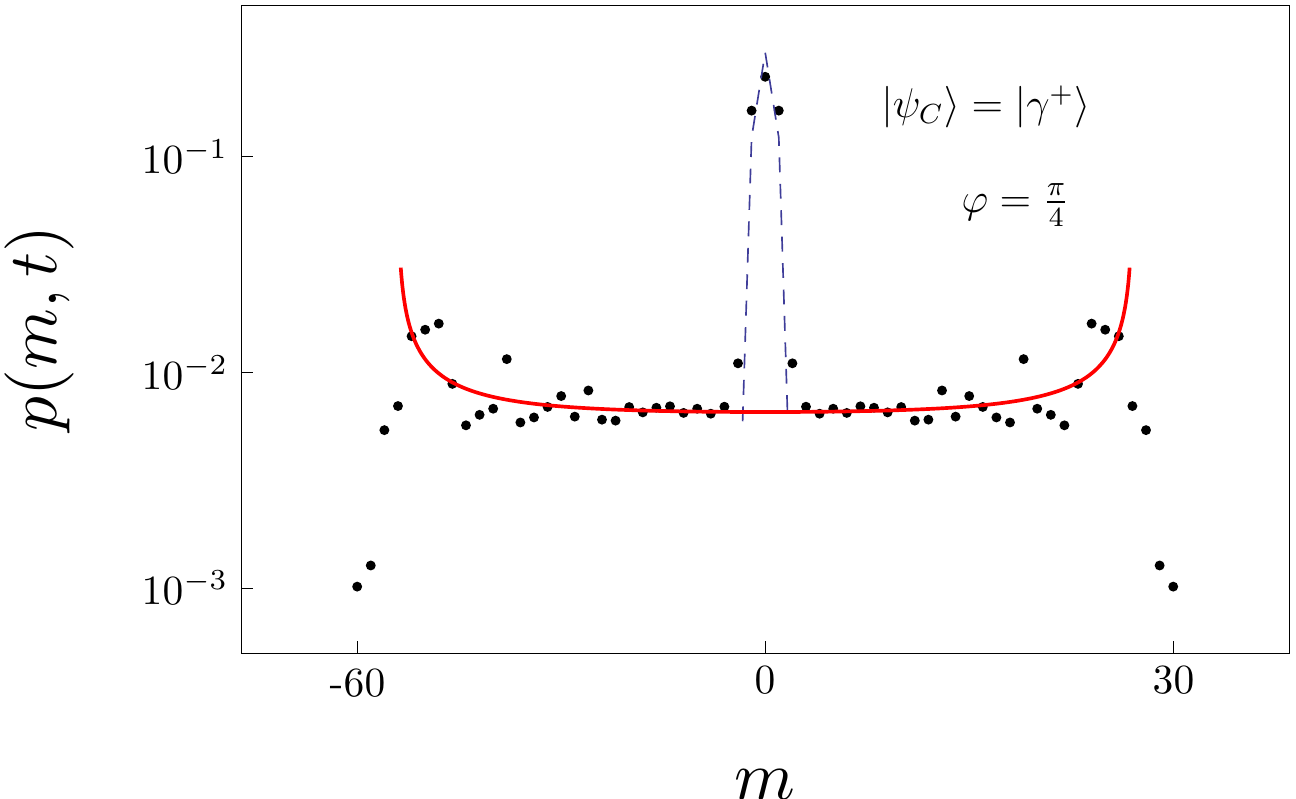}
\caption{(Color
  online) Probability distribution of the three-state walk after $t=100$ steps on a logarithmic scale. The red curve denotes the density (\ref{dist:phi}) and the blue-dashed line corresponds to the localization probability (\ref{loc:phi}). As the initial state we have chosen the coin eigenstate $|\gamma^+\rangle$. For small values of the coin parameter, such as $\varphi=0.01$ depicted in the upper plot, the density bends to zero for $v$ approaching $\pm\eta$. However, for any $\varphi>0$ the density diverges as these points. For increasing values of $\varphi$ the density obtains the inverted-bell shape common for quantum walks. This is illustrated in the lower plot where we choose the coin parameter $\varphi=\frac{\pi}{4}$.}
\label{fig:phi:2}
\end{figure}

Let us first consider the eigenstate $|\gamma^+\rangle$. In the special case of $\varphi=0$ the distribution will not have peaks at the edges, see Fig.~\ref{fig1} for comparison. For very small values of the coin parameter, such as $\varphi=0.01$ illustrated in the upper plot of Fig.~\ref{fig:phi:2}, the density bends down as $v$ approaches $\pm \eta$. Nevertheless, the density depicted by the red curve does not converge to zero. Instead, it diverges at $v=\pm\eta$ for any non-zero value of the coin parameter $\varphi$. With increasing value $\varphi$ the bending of the density diminish and it attains the familiar inverted-bell shape. This is illustrated in the lower plot of Fig.~\ref{fig:phi:2} where we choose the coin parameter $\varphi=\frac{\pi}{4}$.


\begin{figure}[h]
\includegraphics[width=0.45\textwidth]{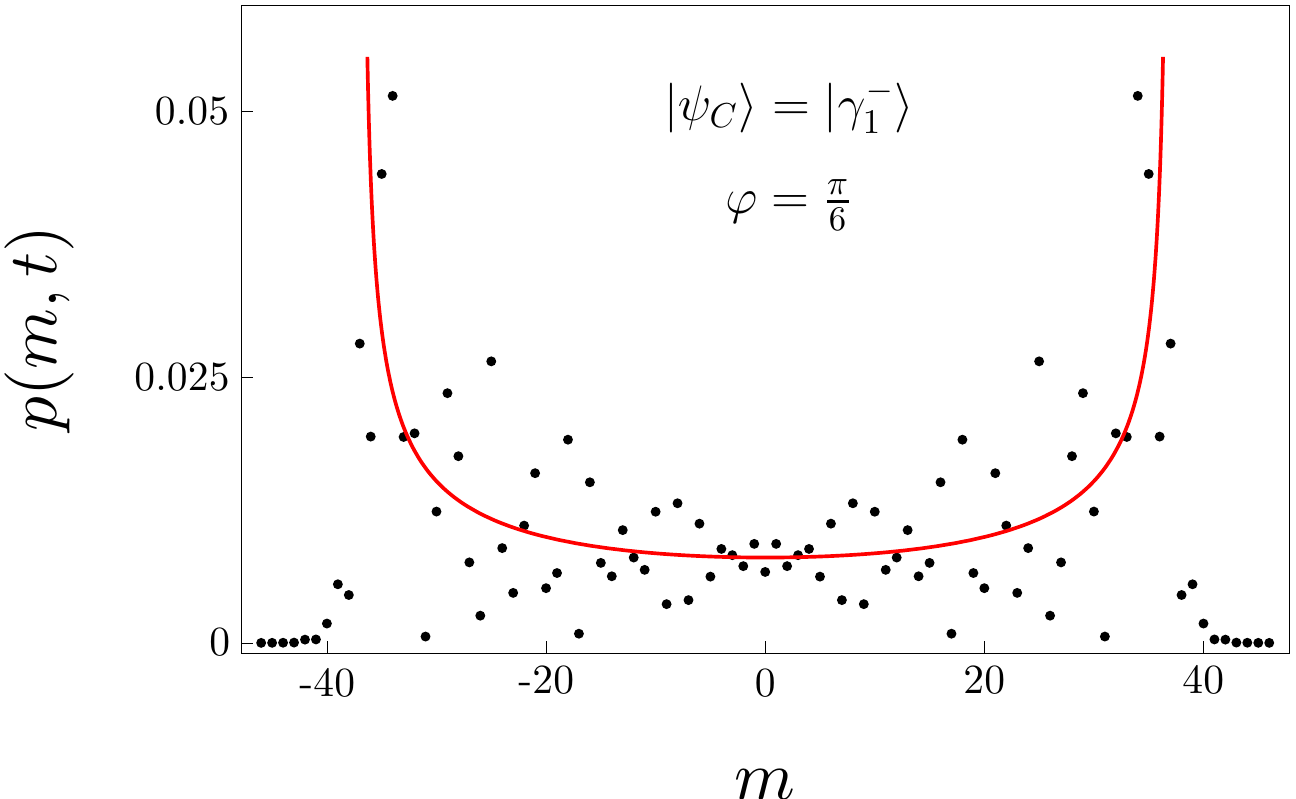}
\caption{(Color
  online) Probability distribution of the three-state walk with the coin parameter $\varphi=\frac{\pi}{6}$ after $t=100$ steps. The red curve corresponds to the density (\ref{dist:phi}). As the initial state we have chosen the coin eigenstate $|\gamma_1^-\rangle$. For this initial coin state the localization effect disappears for all values of $\varphi$.}
\label{fig:phi:3}
\end{figure}

In Fig.~\ref{fig:phi:3} we illustrate the probability distribution after 100 steps for the initial coin state $|\gamma_1^-\rangle$. The coin parameter $\varphi$ was chosen as $\frac{\pi}{6}$. As we have discussed before, choosing this eigenvector as the initial coin state results in the absence of localization for all values of the coin parameter $\varphi$.


\begin{figure}[htbp]
\includegraphics[width=0.45\textwidth]{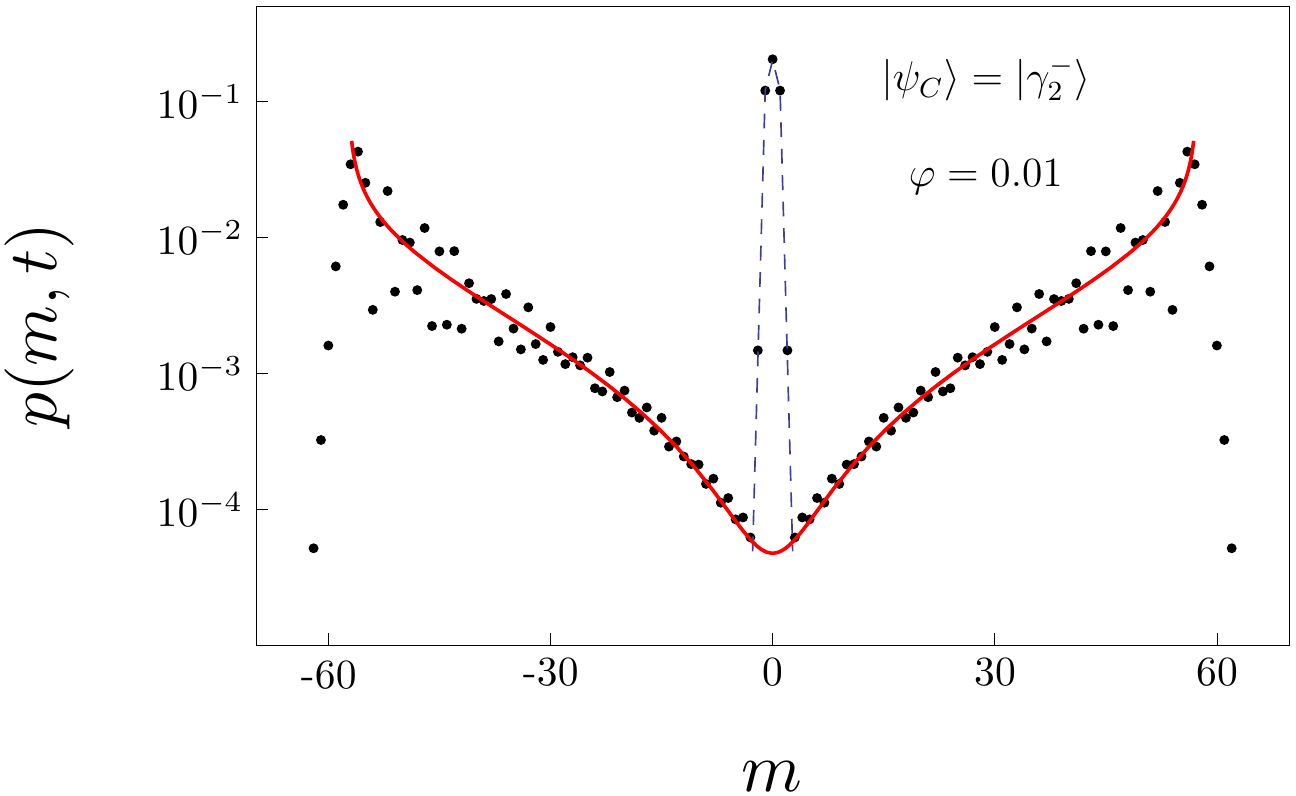}\vspace{12pt}
\includegraphics[width=0.45\textwidth]{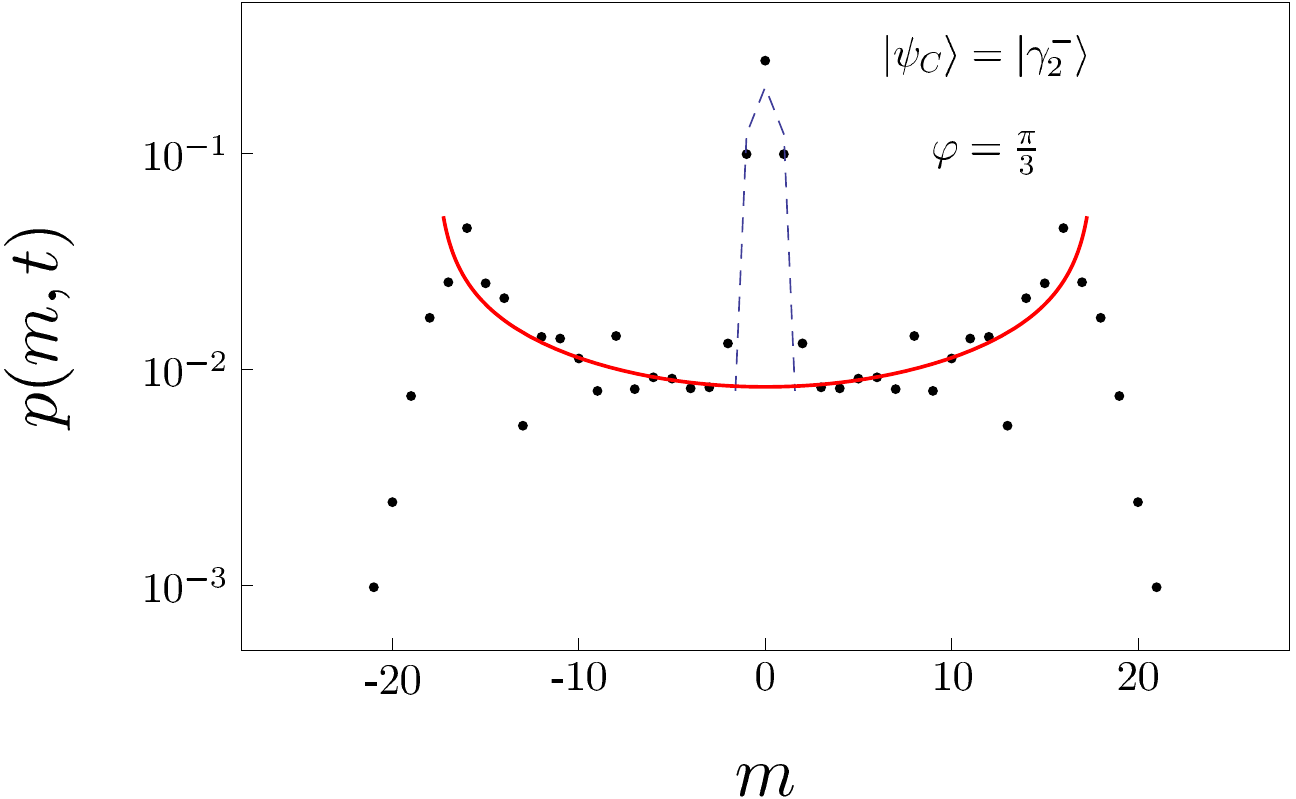}
\caption{(Color
  online) Probability distribution of the three-state walk after $t=100$ steps on a logarithmic scale. The red curve denotes the density (\ref{dist:phi}) and the blue-dashed line corresponds to the localization probability (\ref{loc:phi}). As the initial state we have chosen the coin eigenstate $|\gamma_2^-\rangle$. For small values of the coin parameter the density has a dip around the origin. This is illustrated in the upper plot where we choose $\varphi=0.01$. With increasing values of the coin parameter the density flattens, as depicted in the lower plot for $\varphi=\frac{\pi}{3}$.}
\label{fig:phi:4}
\end{figure}

As a last scenario we consider the eigenstate $|\gamma_2^-\rangle$. In the particular case of $\varphi=0$ the distribution will resemble the one illustrated in Fig.~\ref{fig3}, i.e. the density tends to zero at the origin. For small values of the coin parameter, such as $\varphi=0.01$ depicted in the upper plot of Fig.~\ref{fig:phi:4}, the density maintains a significant dip at the origin. However, it does not converge to zero for any positive $\varphi$. With increasing values of the coin parameter the dip at the origin diminish. This is illustrated in the lower plot of Fig.~\ref{fig:phi:4} where we choose the coin parameter $\varphi=\frac{\pi}{3}$.


To conclude this Section, we have found that the transformation from the standard basis of the coin space to the eigenstate basis is fruitful also for the quantum walks with coin operators (\ref{coin:phi}). In contrast to the results obtained in the previous Section, the present family does not preserve all features of the three-state Grover walk. The exception is the behaviour of localization which is independent $\varphi$.


\section{Conclusions}
\label{sec4}

The limit distributions of two families of three-state quantum walks closely related to the Grover walk have been derived. The first family of quantum walks we have analyzed maintains all properties of the three-state Grover walk. The coin parameter scales the spreading of the walk through the lattice. In contrast, for the second family of quantum walks the features of the three-state Grover walk gradually diminish as the coin parameter increases. The only exception is the behaviour of localization which is the same for all quantum walks within this family.

We have found that limit distributions of both families of quantum walks obtain a particularly simple form when we express the initial coin state in terms of the eigenvectors of the coin operator. This allowed us to reveal the extremal regimes of quantum walk dynamics. Moreover, the dependence of the moments of the distribution on the initial condition can be analyzed in full detail. We have shown that the even moments and the localization probability at the origin depend only on the probabilities of finding the particle initially in the eigenstates of the coin operator. Hence, an incoherent mixture of eigenstates yields in this respect the same results as a pure initial coin state, provided that the aforementioned probabilities are the same. On the other hand, the odd moments and the localization probability outside the origin are determined by the coherence of the initial coin state.

Changing the basis of the coin space to the one formed by the eigenvectors of the coin operator proved to be a very useful tool for the analysis of both families of quantum walks. We expect that this approach will be fruitful also in other models of quantum walks and that it will allow us to uncover otherwise hidden features.

\begin{acknowledgments}

We appreciate the financial support from RVO~68407700. M\v S is grateful for the grant GA\v CR 14-02901P. IB and IJ are thankful for the financial support from SGS13/217/OHK4/3T/14 and GA\v CR 13-33906S.

\end{acknowledgments}


\appendix

\begin{widetext}

\section{Fourier analysis of the eigenvalue family of three-state quantum walks}
\label{app:a}

In this appendix we derive the limit distribution of the three-state quantum walk with the coin (\ref{coin:phi}). In the Fourier representation \cite{ambainis} the evolution operator is given by
$$
\widetilde{U}(k) = {\rm Diag}\left(e^{-i k},1,e^{i k}\right)\cdot C(\varphi) =
\frac{1}{6}\left(
\begin{array}{ccc}
 -e^{-i k}(1+e^{2i\varphi}) & 2e^{-i k}(1+e^{2i\varphi}) & e^{-i k}(5-e^{2i\varphi}) \\
  2(1+e^{2i\varphi}) & 2(1-2e^{2i\varphi}) & 2(1+e^{2i\varphi}) \\
   e^{i k}(5-e^{2i\varphi}) & 2e^{i k}(1+e^{2i\varphi}) & -e^{i k}(1+e^{2i\varphi}) \\
   \end{array}
   \right).
$$
The eigenvalues of this matrix are
$$
\lambda_1 = 1,\quad \lambda_{2,3} = e^{i(\varphi \pm \omega(k))},
$$
where $\omega(k)$ is determined by the dispersion relations
$$
\omega(k) = - \arccos\left(-\frac{1}{3}(2+\cos{k})\cos\varphi\right).
$$
The corresponding eigenvectors have the form
\begin{equation}
\label{eigen:v}
v_1(k) = \sqrt{\frac{2}{5+\cos{k}}}\left(
                 \begin{array}{c}
                   1 \\
                   \frac{1}{2}\left(1+e^{i k}\right) \\
                   e^{i k} \\
                 \end{array}
               \right),\quad v_{2,3}(k) = \frac{1}{\sqrt{n_{2,3}(k)}}\left(
                                         \begin{array}{c}
                                           \left(e^{-i k}+e^{-i(\varphi\pm\omega)}\right)\cos\varphi \\
                                           \cos\omega + e^{\pm i \omega} - e^{-i(2\varphi\pm\omega)} +\cos{k}\cos\varphi \\
                                           \left(e^{-i k}+e^{i(\varphi\pm\omega)}\right)\cos\varphi \\
                                         \end{array}
                                       \right),
\end{equation}
where we have denoted by $n_{2,3}(k)$ the normalization factors which reads
\begin{eqnarray}
\nonumber n_{2,3}(k) & = & \frac{4}{3}\cos^2\varphi\left\{9-4\cos^2\varphi \pm 2\sin\varphi\sqrt{9-(2+\cos{k})^2\cos^2\varphi}-\right. \\
\nonumber & & \left.-\cos{k}\left((4+\cos{k})\cos^2\varphi\pm \sin\varphi\sqrt{9-(2+\cos{k})^2\cos^2\varphi}\right)\right\}.
\end{eqnarray}

\end{widetext}

We note that the first eigenvalue is independent of $k$. Hence, the eigenvector $v_1$ is a stationary state. As we will see later this eigenstate will determine the localization probability. The remaining eigenstates $v_{2,3}$ will contribute to the group-velocity density.

Let us begin with the localization probability that the particle will be found at position $m$ in the asymptotic limit $t\rightarrow + \infty$. We denote by $\psi_C$ the Fourier transformation of the initial state of the quantum walk $|0\rangle\otimes|\psi_C\rangle$. By $f_j(k)$ we mark the overlaps between the eigenstates $v_j(k)$ and $\psi_C$ , i.e.
$$
f_j(k) = (v_j(k),\psi_C).
$$
The probability amplitude of the particle being at position $m$ after $t$ steps of the walk is then given by
\begin{eqnarray}
\label{psi:t}
\psi(m,t) & = & \int\limits_0^{2\pi} \frac{dk}{2\pi}\ e^{-imk} f_1(k)\ v_1(k) + \\
\nonumber & & + \int\limits_0^{2\pi} \frac{dk}{2\pi}\ e^{-imk} e^{i(\varphi+\omega(k)) t} f_2(k)\ v_2(k) +\\
\nonumber & & +\int\limits_0^{2\pi} \frac{dk}{2\pi}\ e^{-imk} e^{i(\varphi-\omega(k)) t} f_3(k)\ v_3(k).
\end{eqnarray}
It follows from the Riemann-Lebesque lemma that the two time-dependent integrals in (\ref{psi:t}) vanish as $t$ approaches infinity. Hence, in the limit $t\rightarrow + \infty$ only the first term in (\ref{psi:t}) remains and we find
$$
\psi_\infty(m) \equiv \lim\limits_{t\rightarrow + \infty} \psi(m,t) = \int\limits_0^{2\pi} \frac{dk}{2\pi}\ e^{-imk} f_1(k)\ v_1(k).
$$
The localization probability $p_\infty(m)$ that the particle is trapped at position $m$ as $t$ approaches infinity is then given by $|\psi_\infty(m)|^2$. We see that it depends solely on the stationary state $v_1(k)$. Since $v_1(k)$ is independent of the coin parameter $\varphi$ the localization probability $p_\infty(m)$ is the same for the whole family of quantum walks. The result coincides with the one for the three-state Grover walk corresponding to the value $\varphi=0$.

We now turn to the derivation of the group-velocity density. This can be deduced by calculating the moments of the particle's position in the Fourier representation \cite{watabe}. Let $m$ denote the position of the particle after $t$ steps of the quantum walk. One can show \cite{Grimmett} that the $n$-th moment of the re-scaled position $\frac{m}{t}$ converges in the limit of $t\rightarrow +\infty$. Following the approach of \cite{watabe} we find that the limiting value of the moment is given by
\begin{widetext}
\begin{equation}
\label{grover:moment}
\lim\limits_{t\rightarrow +\infty}\left\langle \left(\frac{m}{t}\right)^{n}\right\rangle = \int\limits_0^{2\pi} \left(\frac{d\omega}{dk}\right)^n\ \left((-1)^n \left|f_2(k)\right|^2 + \left|f_3(k)\right|^2\right)\ \frac{dk}{2\pi}.
\end{equation}
\end{widetext}
Let us determine the overlap functions $f_j(k)$, respectively the term $\left((-1)^n \left|f_2(k)\right|^2 + \left|f_3(k)\right|^2\right)$. We note that
\begin{equation}
\label{group:v}
v = \frac{d\omega}{dk} = \frac{\cos\varphi\sin{k}}{\sqrt{9-(2+\cos{k})^2\cos^2\varphi}},
\end{equation}
is an odd function of $k$. Hence, for even $n$ it is sufficient to find only the even part of $|f_2(k)|^2+|f_3(k)|^2$, since the contribution of the odd part to the integral (\ref{grover:moment}) is zero. Similarly, for odd $n$ it is sufficient to determine the odd part of $-|f_2(k)|^2+|f_3(k)|^2$. To do this we consider the decomposition of the initial coin state $|\psi_C\rangle$ in the eigenstate basis as in (\ref{init:phi}). The Fourier transformation of the initial state of the walk $|0\rangle\otimes|\psi_C\rangle$ is then given by
$$
\psi_C = \left(
           \begin{array}{c}
             \frac{g_1}{\sqrt{6}}+\frac{g_2}{\sqrt{2}}+\frac{g_+}{\sqrt{3}} \\
             \frac{g_+}{\sqrt{3}}-\sqrt{\frac{2}{3}} g_1 \\
             \frac{g_1}{\sqrt{6}}-\frac{g_2}{\sqrt{2}}+\frac{g_+}{\sqrt{3}} \\
           \end{array}
         \right)
$$
With the explicit form of the eigenvectors (\ref{eigen:v}) we then find that the even part of $|f_2(k)|^2+|f_3(k)|^2$ reads
\begin{widetext}
$$
\left\{|f_1(k)|^2 + |f_2(k)|^2\frac{}{}\right\}_{{\rm even}} = 3|g_1|^2 + 5|g_2|^2 - 2 + \left(1 - |g_1|^2 - 2|g_2|^2\right) \frac{12}{5+\cos{k}}
$$
In a similar way, the odd part of $-|f_2(k)|^2+|f_3(k)|^2$ is given by
\begin{eqnarray}
\nonumber \left\{-|f_2(k)|^2 + |f_3(k)|^2\frac{}{}\right\}_{{\rm odd}} & = & \left(-\sqrt{3}(g_1\overline{g_2} + \overline{g_1}g_2 + i(g_1\overline{g_2}-\overline{g_1}g_2)\tan\varphi) +i\sqrt{6}(g_2\overline{g_+}-\overline{g_2}g_+)\frac{2+\cos{k}}{5+\cos{k}}\tan\varphi\right)\times\\ \nonumber & & \times \frac{\cos\varphi\sin{k}}{\sqrt{9-(2+\cos{k})^2\cos^2\varphi}}
\end{eqnarray}
The last part of the derivation of the group-velocity density is to make the substitution $k\rightarrow v$ in the integral (\ref{grover:moment}), where the group-velocity $v$ is defined in (\ref{group:v}). The transformation from $k$ to $v$ is not unique and has to be done separately in two intervals. For $k\in(0,k_0)\cup (2\pi -k_0,2\pi)$, where
$$
k_0 = \arccos\left(\frac{1}{4\cos^2\varphi}(9 - 5 \cos^2\varphi - 3\sin\varphi\sqrt{9-\cos^2\varphi})\right),
$$
we find from (\ref{group:v}) the that the following relations hold
\begin{eqnarray}
\nonumber \cos{k} & = & \frac{2v^2+\sqrt{1+3v^2-9v^2\frac{1-v^2}{\cos^2\varphi}}}{1-v^2},\\
\nonumber \sin{k} & = & v \frac{\sqrt{9(1-v^2)^2-\cos^2\varphi\left(2+\sqrt{1+3v^2-9v^2\frac{1-v^2}{\cos^2\varphi}}\right)^2}}{\cos\varphi(1-v^2)}.
\end{eqnarray}
Similarly, in the interval $k\in (k_0,2\pi-k_0)$ we find the identities
\begin{eqnarray}
\nonumber \cos{k} & = & \frac{2v^2-\sqrt{1+3v^2-9v^2\frac{1-v^2}{\cos^2\varphi}}}{1-v^2},\\
\nonumber \sin{k} & = & v \frac{\sqrt{9(1-v^2)^2-\cos^2\varphi\left(2-\sqrt{1+3v^2-9v^2\frac{1-v^2}{\cos^2\varphi}}\right)^2}}{\cos\varphi(1-v^2)}.
\end{eqnarray}
Performing all steps of the substitution from $k$ to $v$ is quite tedious but after some algebra we find that the moments (\ref{grover:moment}) can be re-written in the form
$$
\lim\limits_{t\rightarrow +\infty}\left\langle \left(\frac{m}{t}\right)^{n}\right\rangle = \int\limits_{-\eta}^{\eta} v^n\ w(v)\ dv,
$$
where $\eta$ is the maximum of the group velocity $v(k)$ which is achieved at $k=k_0$. We find that it is given by
$$
\eta = \frac{1}{\sqrt{6}}\sqrt{3-\cos^2{\varphi}-\sin{\varphi}\sqrt{9-\cos^2{\varphi}}}.
$$
By $w(v)$ we have labeled the group-velocity density which can be expressed in the form
\begin{eqnarray}
\nonumber w(v) = \frac{1}{6\pi (1-v^2) \Theta} & & \left[\frac{}{}\left(3|g_1|^2 + 5|g_2|^2 - 2\right)\Lambda_+ + \left(1 - |g_1|^2 - 2|g_2|^2\right) \Omega - \right.\\
\nonumber & & \left. -  \sqrt{3}v(g_1\overline{g_2} + \overline{g_1}g_2 + i(g_1\overline{g_2} -\overline{g_1}g_2)\tan\varphi)\Lambda_+ +  i v(g_2\overline{g_+}-\overline{g_2}g_+) \Xi\frac{}{}\right].
\end{eqnarray}
Here we have used the notation
\begin{eqnarray}
\nonumber \Lambda_\pm & = & \Phi_+ \pm \Phi_-,\quad  \Phi_\pm = \sqrt{9(1-v^2) - (5+3v^2)\cos^2\varphi \pm 12 \Theta \cos\varphi },\\
\nonumber \Omega & = & 4\cos\varphi \frac{(5-3v^2)\cos\varphi \Lambda_+ + 3\Theta \Lambda_-}{8\cos^2\varphi + 3v^2\sin^2\varphi},\\
\nonumber \Xi & = & 3\sqrt{6}\tan\varphi \frac{(v^2 + \cos^2\varphi)\Lambda_+ -  \Theta\cos\varphi \Lambda_-}{8\cos^2\varphi + 3v^2\sin^2\varphi}, \\
\nonumber \Theta & = & \sqrt{(\eta^2-v^2)\left(\eta^2-v^2+\sin\varphi\sqrt{1-\frac{\cos^2\varphi}{9}}\right)}.
\end{eqnarray}

\end{widetext}



\begin{thebibliography}{x}
\bibitem{machida:note}
We note that Machida in \cite{machida} uses a different parametrization of the coin operator (\ref{coin:rho}). Our parameter $\rho$ corresponds to the angle $\theta$ used in \cite{machida} through $\cos\theta = 2\rho^2-1$ and $\sin\theta = 2\rho\sqrt{1-\rho^2}$. We prefer to use the parametrization (\ref{coin:rho}) as it was originaly introduced in \cite{stef:cont:def} since $\rho$ has a direct dynamical interpretation as a peak velocity.
\end{thebibliography}
\end{document}